\documentclass[a4paper,
               keeplastbox,   
               ]{jacow}
\usepackage{pdfpages,multirow,ragged2e} %
\makeatletter%
	\ifboolexpr{bool{xetex}}
	 {\renewcommand{\Gin@extensions}{.pdf,%
	                    .png,.jpg,.bmp,.pict,.tif,.psd,.mac,.sga,.tga,.gif,%
	                    .eps,.ps,%
	                    }}{}
\makeatother

\ifboolexpr{bool{xetex} or bool{luatex}} 
 {}                                      
 {\usepackage[utf8]{inputenc}}           

\usepackage[USenglish]{babel}

\ifboolexpr{bool{jacowbiblatex}}%
 {%
  \addbibresource{jacow-test.bib}
  \addbibresource{biblatex-examples.bib}
 }{}
\begin{document}

\title{Update of beam coupling impedance evaluation by the stretched-wire method}

\author{ T. Toyama\thanks{email address: takeshi.toyama@kek.jp}, A. Kobayashi, T. Nakamura, M. Yoshii, C. Ohmori, K. Hasegawa, Y. Sugiyama, \\
T. Shibata, K. Ishii, KEK, Tokai, Ibaraki,  Japan  \\
Y. Shobuda, F. Tamura, JAEA, Tokai, Ibaraki, Japan \\ 
K. Hanamura, T. Kawachi, Mitsubishi Electric System \& Service Co.,Ltd, Tsukuba, Ibaraki, Japan }
	
\maketitle

\begin{abstract}
   In many cases beam coupling impedances or wake fields are calculated with computer simulator 
such as CST Studio Suite, GdfidL Electromagnetic Field Simulator and so on.
But evaluation with the stretched-wire method is still very useful by its flexibility
to the change of the device-under-test configuration, speed to get results and, 
more than anything, its accessibility on the real devices.
One of the drawbacks in the practical procedure, difficulty of Thru-Reflect-Line calibration,
is overcome using the calibration method recently introduced in high frequency vector network analyzers, 
"2-X THRU de-embedding". Here the procedure is explained in detail.
The method has been successfully applied to RF cavity and FX kicker measurement in the J-PARC MR.
\end{abstract}

\section{INTRODUCTION}
In recent operations of the J-PARC MR\cite{Igarashi},
longitudinal microstructure in the debunching bunches causes a problem 
including beam instabilities at the flattop energy in the slow extraction (SX) mode. 
In the fast extraction (FX) mode a similar problem is foreseen, 
because longitudinal phase space manipulation will be required 
for peak beam current reduction and debunching is included in this process. 
As a first step for understanding this longitudinal microwave instability and taking a countermeasure, 
we are evaluating the longitudinal coupling impedances of major devices of a large coupling impedance such as kickers, septa, RF cavities. 

The stretched wire method is a widely used coupling impedance measurement method of the actual devices.
The procedure was rather complicated for a rigorous measurement method which needs Thru-Reflect-Line (TRL) calibration \cite{Walling}. 
A simplified method by measuring only a device under test (DUT), omitting other measurements and using numerical correction of phase shift $ \theta = k \ell $ 
cannot cancel the mismatch effect at the transition parts between 50 $ \Omega $ coaxial cables. 
This drawback is overcome using the calibration method recently introduced in high frequency vector network analyzers, "2-X THRU de-embedding". 
The procedure of this method is described 
and the present status of the mapping from $ S_{21} $ to $ Z_L $ using conventional methods is reported.

\section{PROCEDURE OF "2X-THRU de-embedding"}
\begin{figure}[!htb]
   \centering
   \includegraphics*[width=.45\columnwidth]{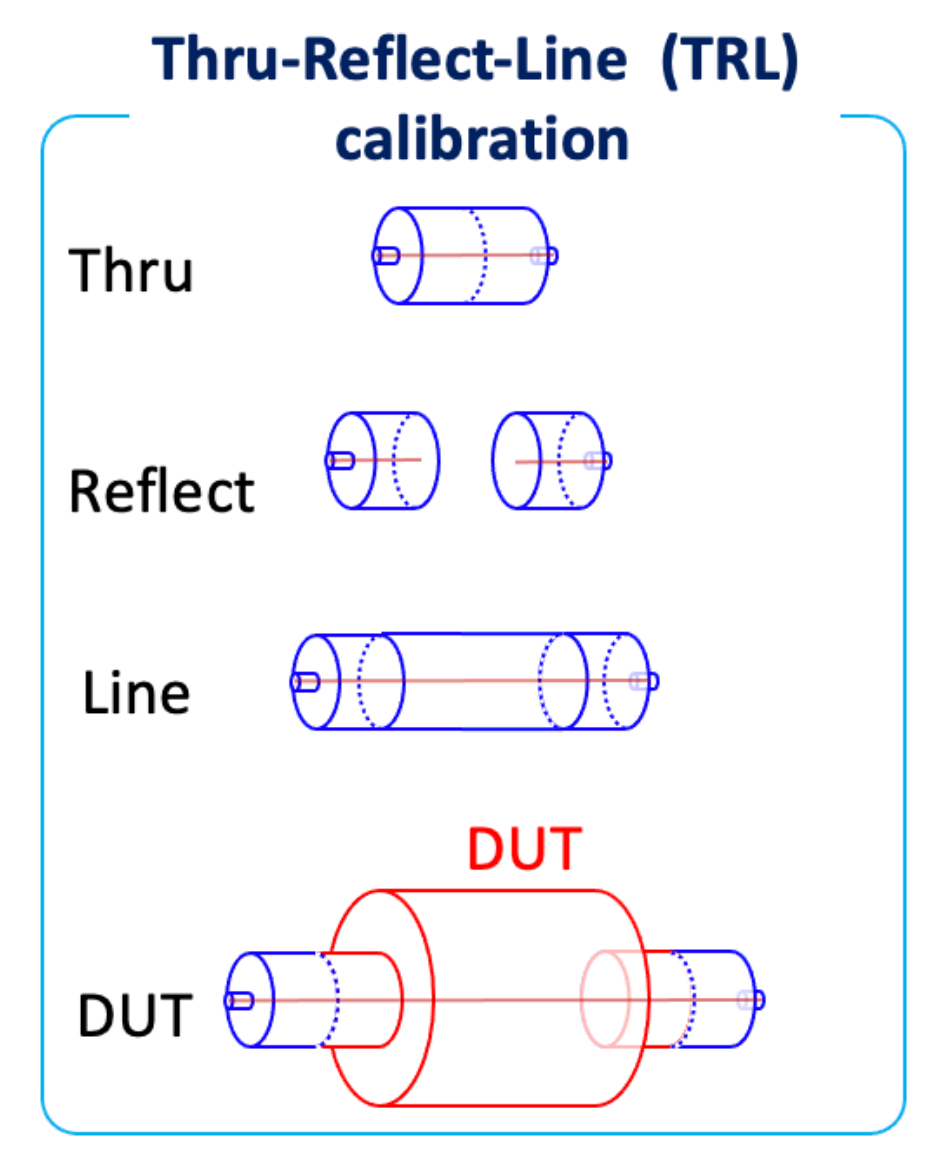}
   \includegraphics*[width=.45\columnwidth]{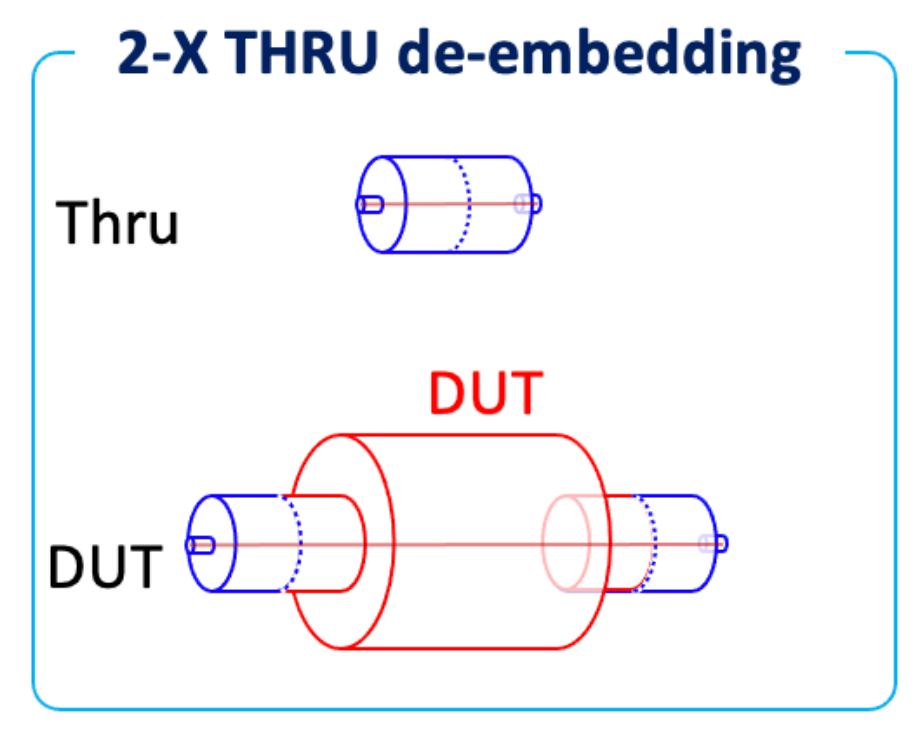}
   \caption{Two calibration methods. The left figure is the setup variation for the TRL method.
   The right figure is the setup variation for the 2X-THRU de-embedding method.}
   \label{fig:CALmethods}
\end{figure}

Setups for TRL and 2-X THRU de-embedding methods are summarized in Fig.\ref{fig:CALmethods}.
The left figure shows the setup variation for the TRL method.
"Thru", "Reflect", "Line" and "DUT" fixtures and measurements are necessary. 
   The right figure shows the setup variation for the 2X-THRU de-embedding method \cite{Ellison}.
   The setup is much simplified. The time and effort of measurement is also simplified.

\begin{figure}[!htb]
   \centering
   \includegraphics*[width=.55\columnwidth]{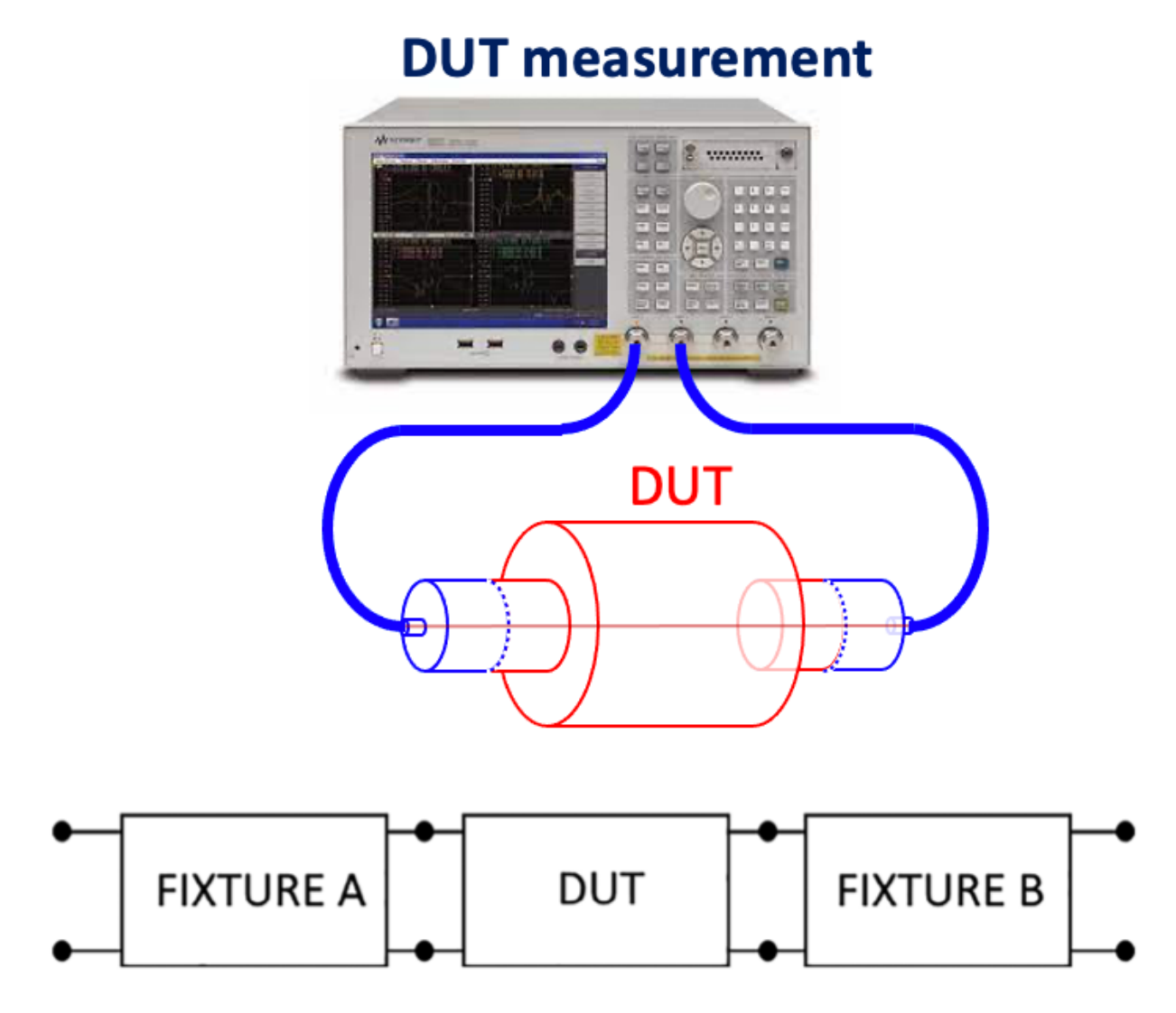}
   \includegraphics*[width=.4\columnwidth]{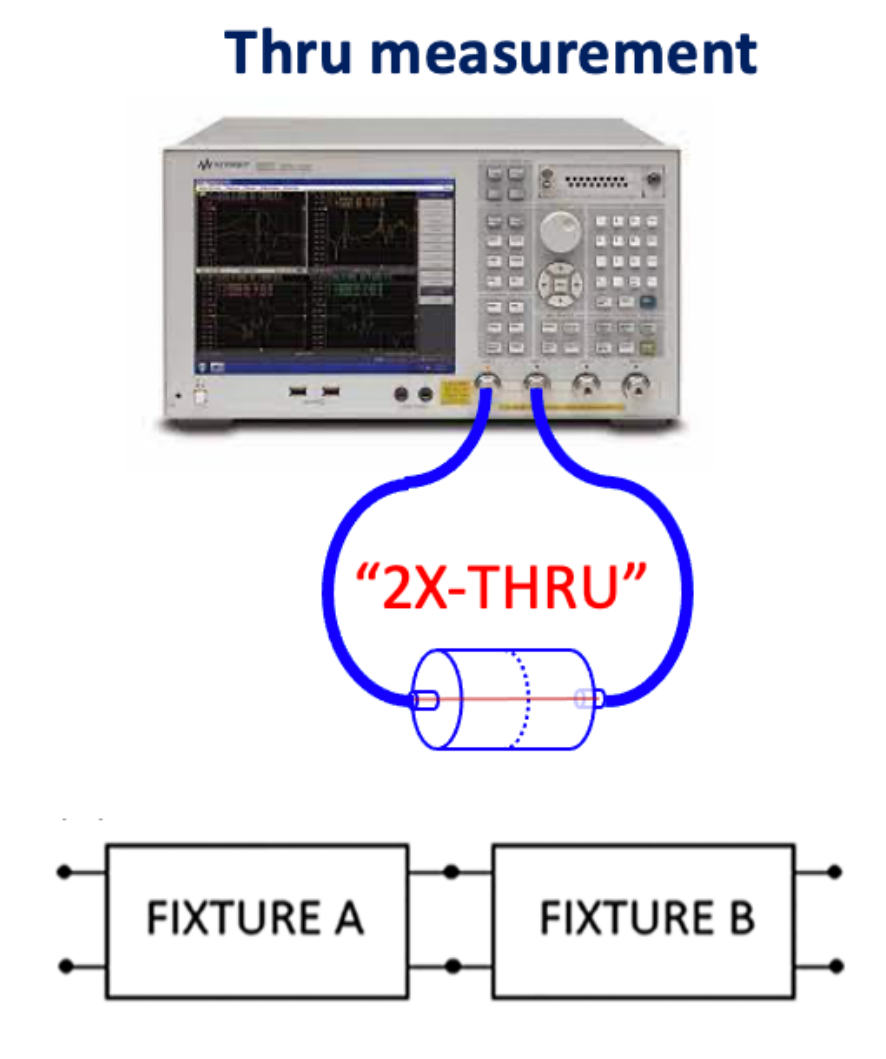}
   \includegraphics*[width=.75\columnwidth]{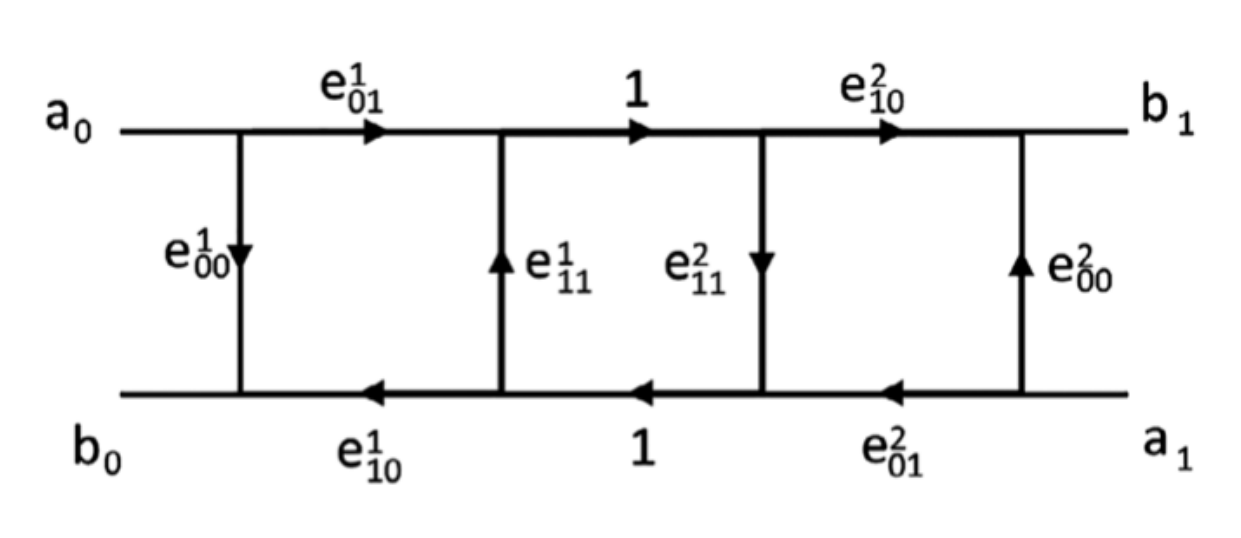}
   \caption{The two setups for 2X-THRU de-embedding. 
   The top left figure is the schematic of a DUT measurement with a corresponding block diagram \cite{Ellison}.
   The top right is for a 2X-THRU measurement.  
   Bottom figure is a signal flow graph of the 2X-THRU measurement \cite{Ellison}. }
   \label{fig:2XTHRU}
\end{figure}

We can extract the DUT S-parameter by removing the effects of the FIXTURE A and B from the whole S-parameter in Fig. \ref{fig:2XTHRU},
by characterizing the S parameters of FIXTURE A and B, separately  \cite{Ellison}.

\begin{figure}[!htb]
   \centering
   \includegraphics*[width=.75\columnwidth]{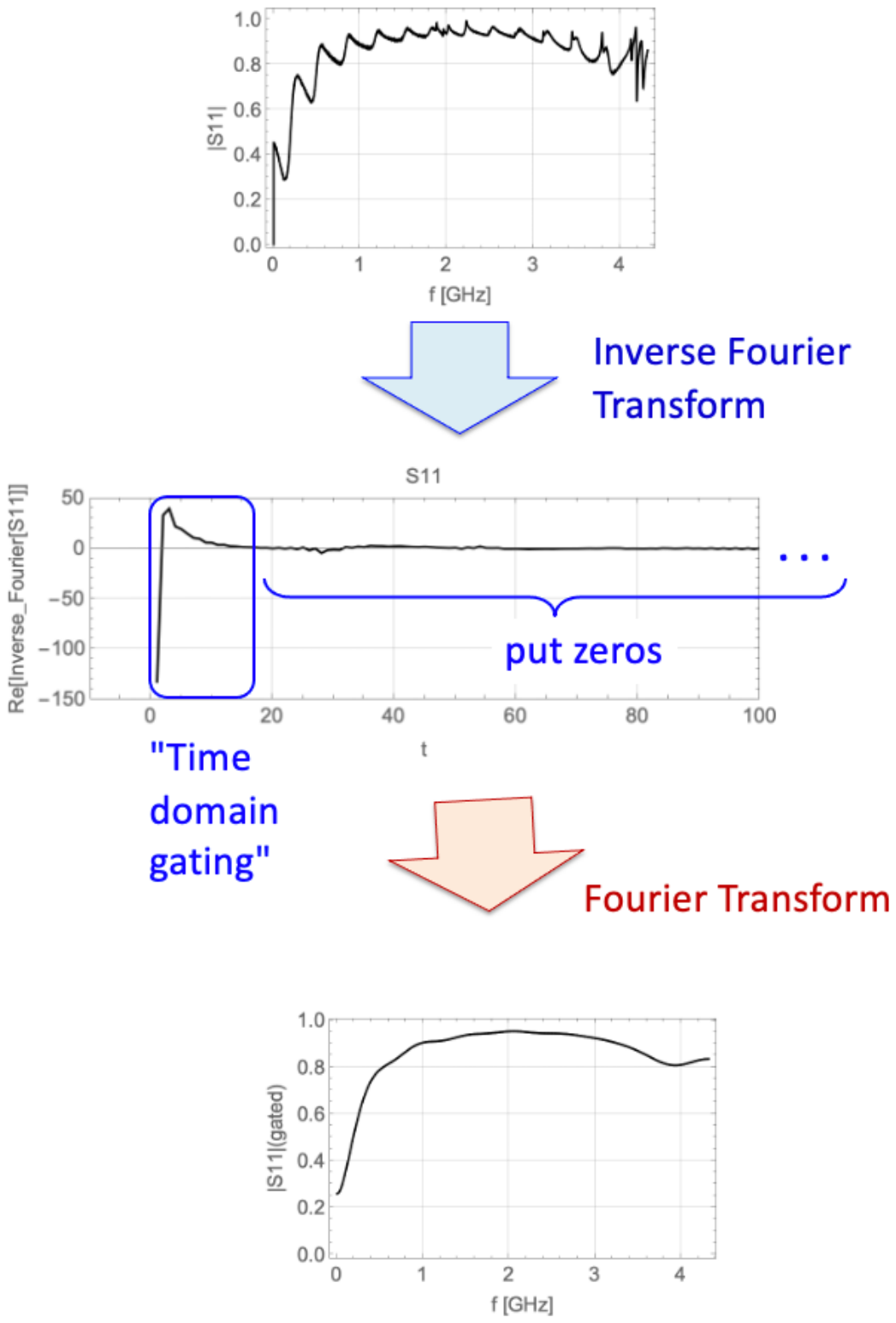}
   \caption{Characterization of the 1X-THRU. 
   Reflection coefficient $S_{11}$ (top) is inverse Fourier transformed, then put zeros in the time period larger than one-way trip in 1X structure (middle). Then Fourier transformed to acquire $e_{00}$ (bottom). }
   \label{fig:1X}
\end{figure}

Originally, there are four equations of S-parameters $S_{ij}$ for eight unknowns $ e^i_{jk} $.
Assuming that $ e^1_{10} = e^1_{01} $,  $ e^2_{10} = e^2_{01} $, 
number of unknowns reduces to six.
Variables $ e^1_{00}  $ and $ e^2_{00} $ are obtained by the procedure
 intuitively summarized in  Fig.\ref{fig:1X}.
 
We can obtain  $ e^i_{jk} $ by solving the equations 
\begin{equation}\label{eq:1}
     e^1_{11} = \frac{ S_{11} - e^1_{00}  }{ S_{21}  },
\end{equation}

\begin{equation}\label{eq:2}
     e^2_{11} = \frac{ S_{22} - e^2_{00}  }{ S_{12}  },
\end{equation}

\begin{equation}\label{eq:3}
      (e^1_{10})^2 = S_{21} ( 1 -  e^1_{11} e^2_{11}  ),
\end{equation}

\begin{equation}\label{eq:4}
       (e^2_{10})^2 = S_{12} ( 1 -  e^1_{11} e^2_{11}  ).
\end{equation}

Then we extract the DUT S-parameter by removing the effects of the FIXTURE A and B. 
At the beginning of this procedure, a scattering matrix should be renormalized from the characteristic impedance of the network analyzer of 50 $\Omega$ to that of the coaxial transmission line consisting of a wire and a circular pipe, $Z_C$, typically $404 \Omega$ for a RF system and $428 \Omega$ for a kicker system.
Finally using $Z_C$ and appropriate formula, obtain the longitudinal coupling impedance, $Z_L(\omega)$.
We tried three formulas, "standard log formula", "improved log formula" and "lumped element formula" \cite{Jensen}

\begin{equation}\label{eq:5}
     Z_{L}(\omega) = - 2 Z_C {\rm log} \left( \frac{ S^{DUT}_{21}  }{ e^{ -jk \ell }  } \right),
\end{equation}

\begin{equation}\label{eq:6}
       Z_{L}(\omega)  =  - 2 Z_C {\rm log} \left( \frac{ S^{DUT}_{21}  }{ e^{ -jk \ell }  } \right) \left[  1 +  \frac{j c}{2 \omega \ell} {\rm log}\left( \frac{ S^{DUT}_{21}  }{ e^{ -jk \ell }  } \right)  \right],
\end{equation}

\begin{equation}\label{eq:7}
        Z_{L}(\omega)  =  \frac{ 2 Z_C (1-S_{21}) }{ S_{21} },
\end{equation}

where $k = \omega / c$, $\omega$ is the angular frequency, $c$ the light velocity and $\ell$ is the length of the DUT.

\section{$Z_L$ of the Accelerating RF system}

\begin{figure}[!htb]
   \centering
   \includegraphics*[width=.75\columnwidth]{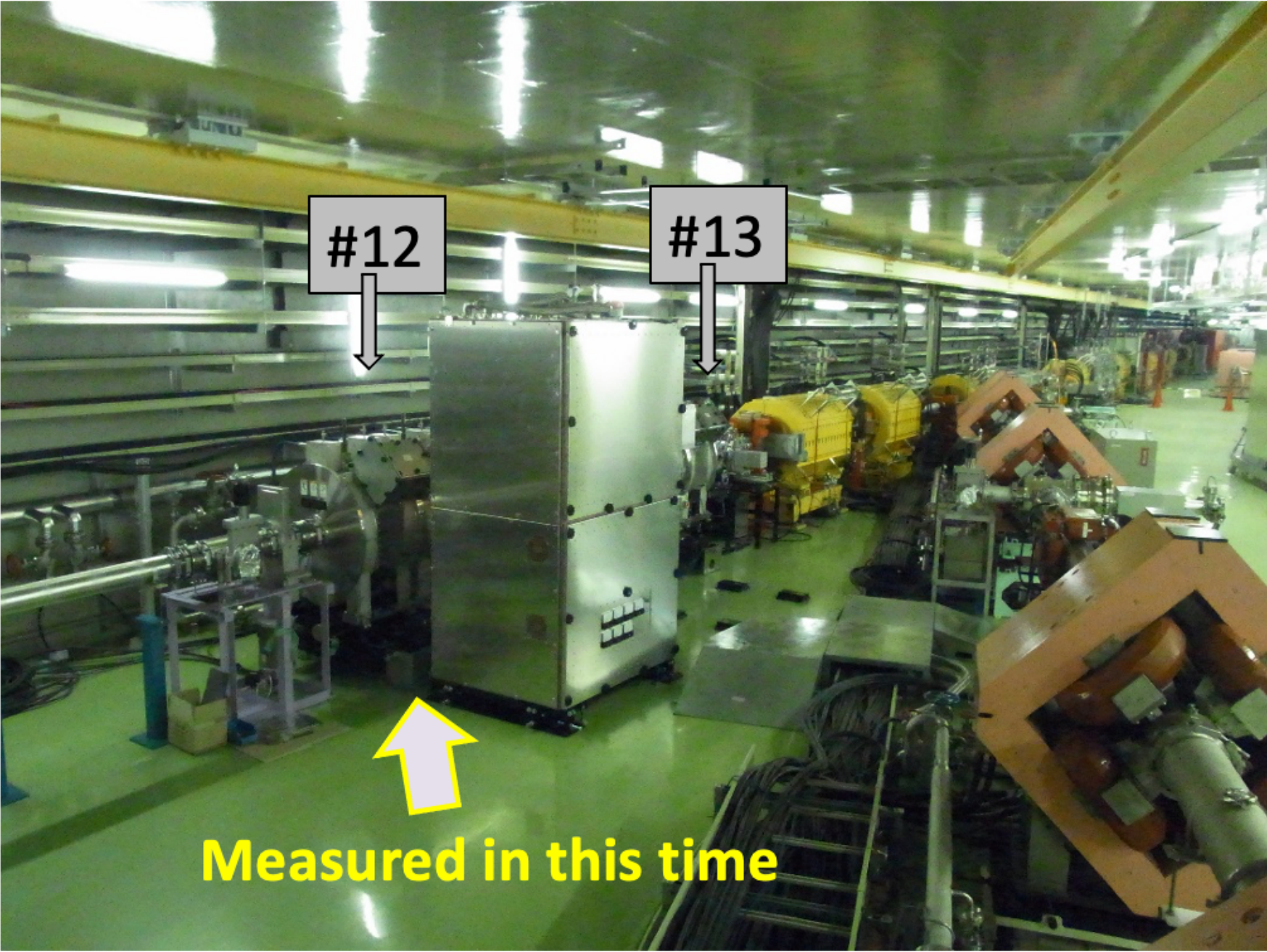}
   \caption{The RF system at upstream of INS-A in the MR.}
   \label{fig:Fig4}
\end{figure}

\begin{figure}[!htb]
   \centering
   \includegraphics*[width=.55\columnwidth]{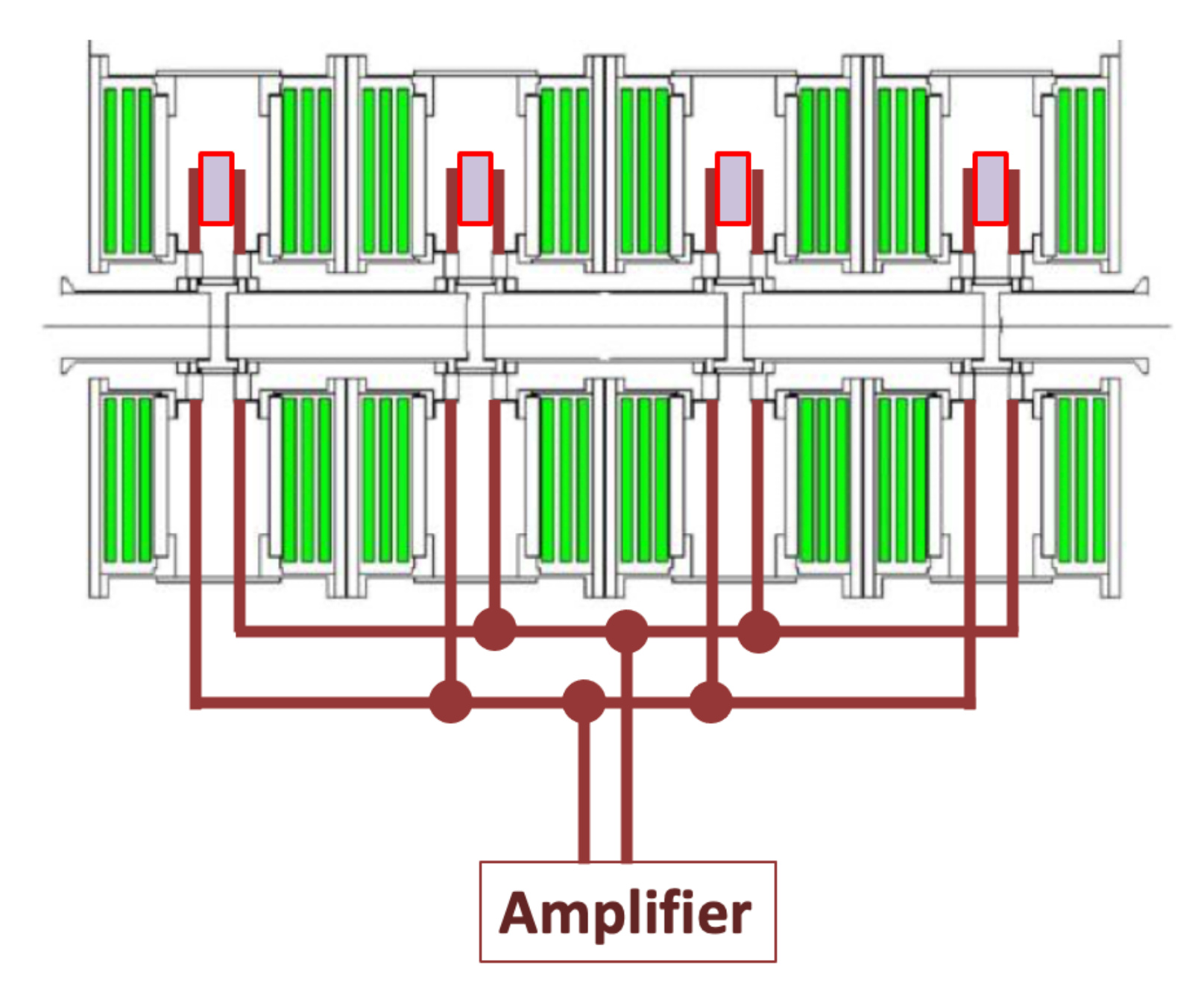}
   \caption{Schematic diagram of the RF system.}
   \label{fig:Fig5}
\end{figure}

Our RF system in the MR has been upgraded according to the plan tabulated in Table~\ref{tab:AccRF} \cite{Igarashi}.
We measured newly installed RF system (cavity and final amplifier) at upstream of INS-A in the shutdown period in Sep. and Oct. 2021 (Fig.\ref{fig:Fig4}).
The setup can be changed between fundamental and 2nd harmonic resonance configuration 
by additional capacitors at the acceleration gaps (Fig.\ref{fig:Fig5}).
We measured several setups including a fundamental, 2nd harmonic, gaps shorted by bus bars and only gaps.

\begin{table}[!hbt]
   \centering
   \caption{Upgrade plan of the RF system in the MR}
   \begin{tabular}{lccc}
       \toprule
       \textbf{} & \textbf{Present}                      & \textbf{2022}                & \textbf{2026} \\
       \midrule
         MR Cycle                         &   2.48 s  &  1.32 s  &  1.16 s   \\ 
          Fundamental Cavities    &     7        &    9 &  11   \\ 
           2nd Harmonic Cavities   &     2       &  2  & 2  \\ 
           Accelerating Voltage      &  300 kV &  510 kV  &  600 kV  \\
           2nd Harmonic Voltage   &   110 kV  &  110 kV  &  110 kV  \\
       \bottomrule
   \end{tabular}
   \label{tab:AccRF}
\end{table}

\begin{figure}[!htb]
   \centering
   \includegraphics*[width=.49\columnwidth]{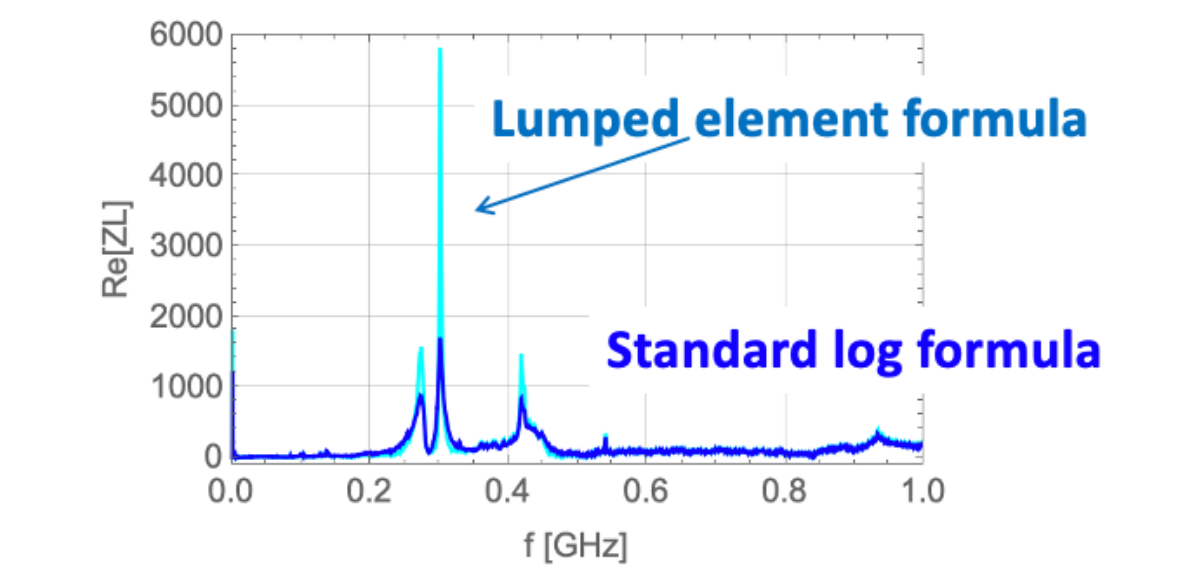}
   \includegraphics*[width=.49\columnwidth]{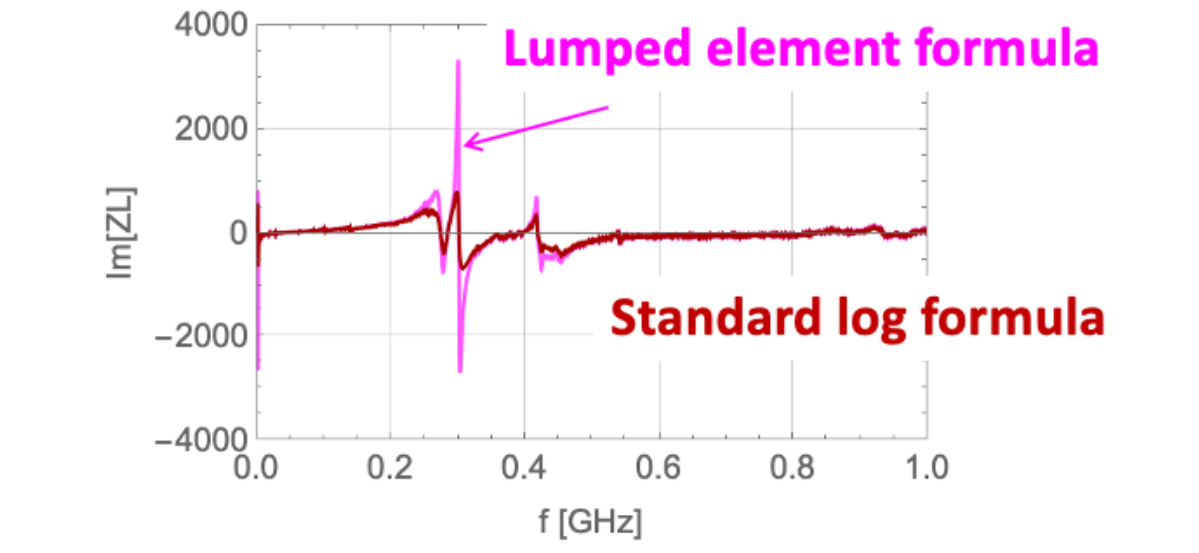}
   \caption{$Z_L$ of the fundamental frequency configuration.}
   \label{fig:Fig6}
\end{figure}

\begin{figure}[!htb]
   \centering
   \includegraphics*[width=.49\columnwidth]{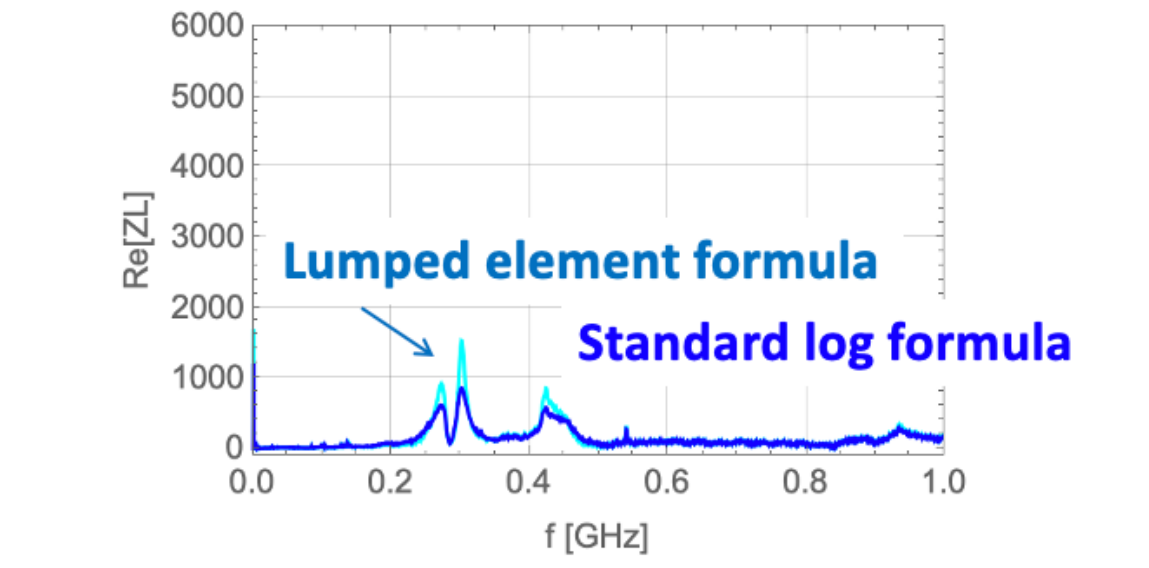}
   \includegraphics*[width=.49\columnwidth]{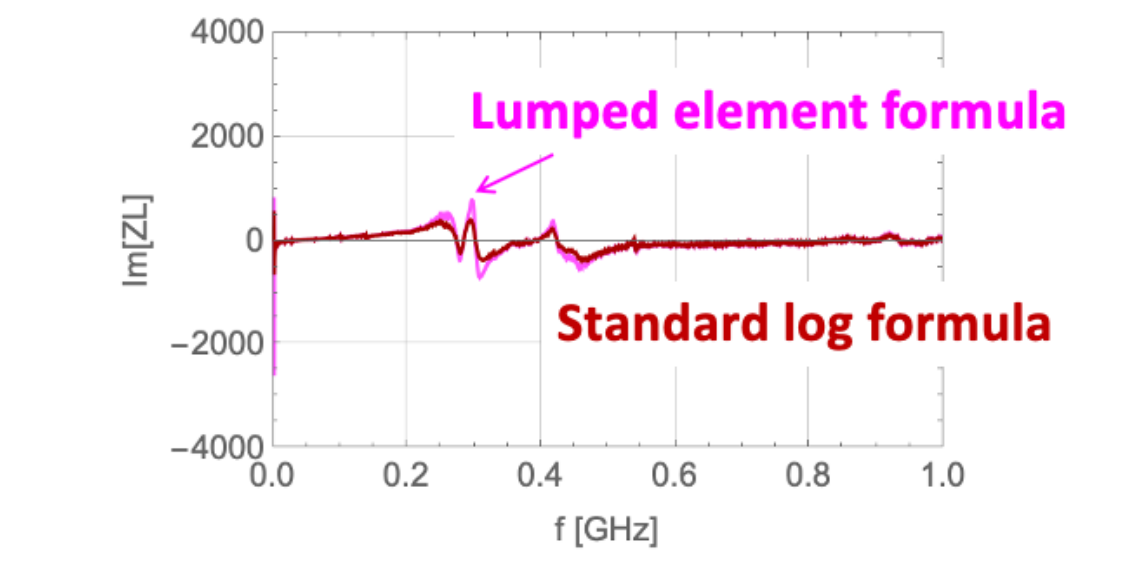}
   \caption{$Z_L$ of the fundamental frequency configuration with a resistive damper.}
   \label{fig:Fig7}
\end{figure}

\begin{figure}[!htb]
   \centering
   \includegraphics*[width=.43\columnwidth]{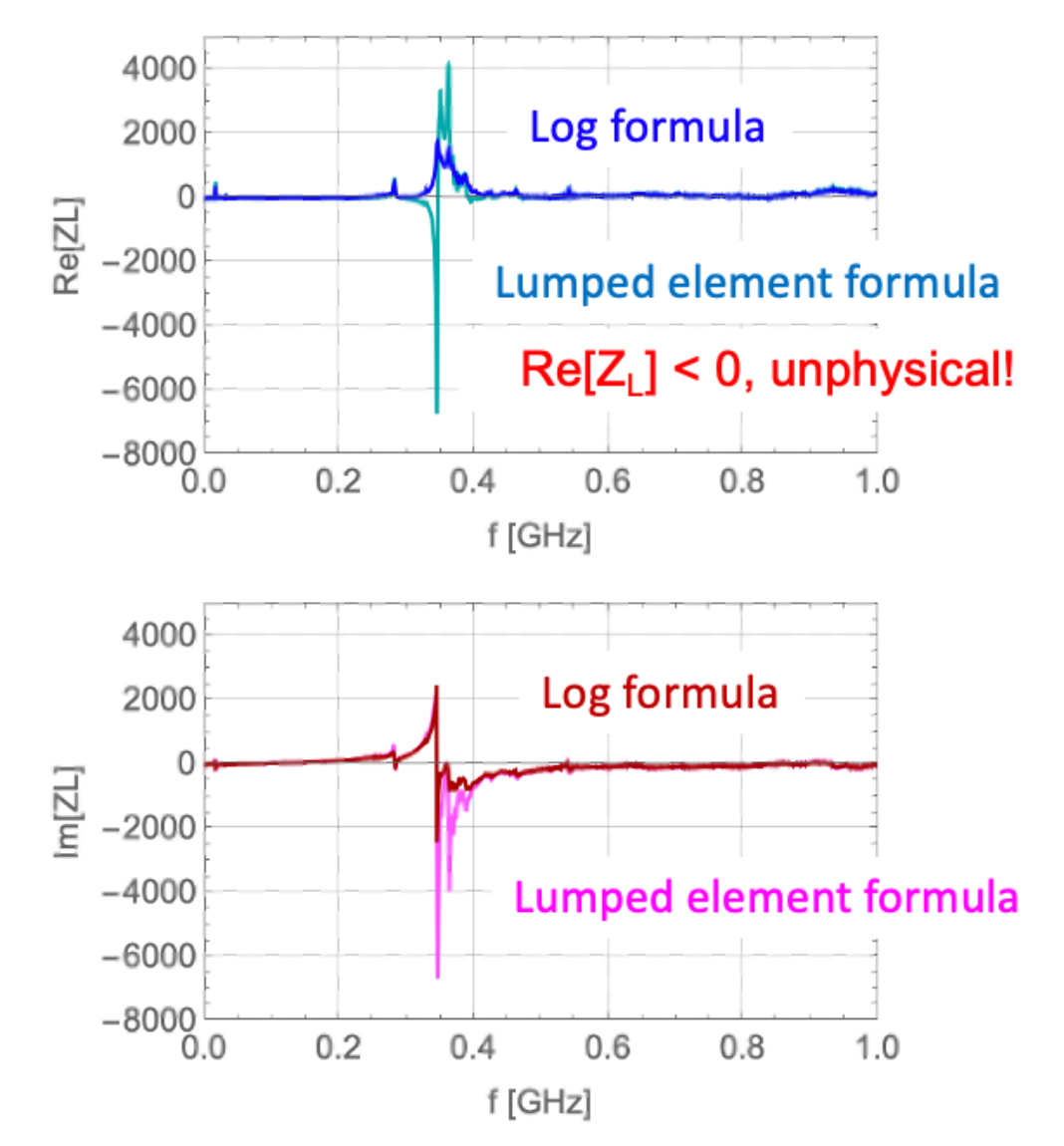}
   \includegraphics*[width=.55\columnwidth]{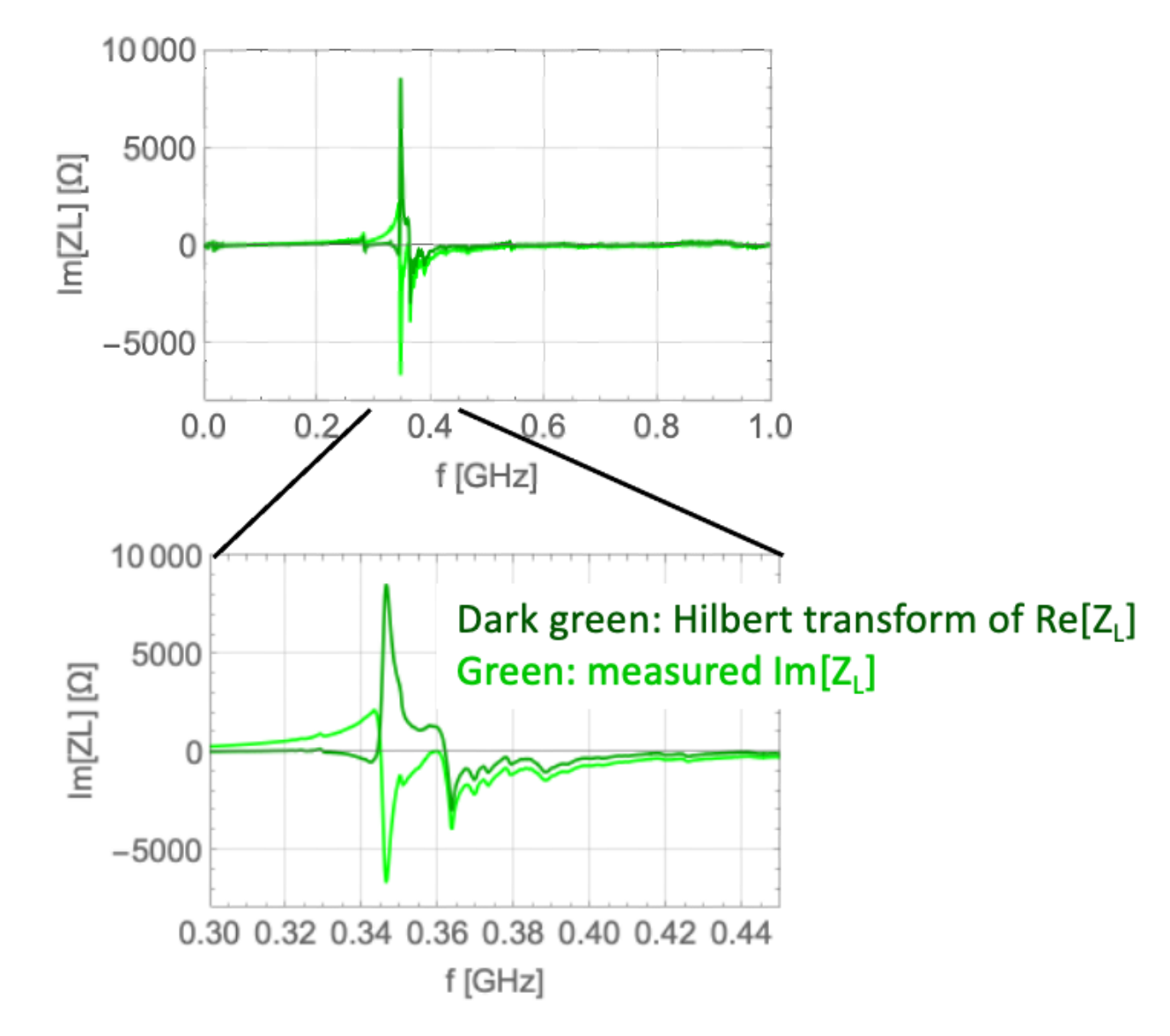}
   \caption{(Left) results calculated by the lumped element formula and the standard log formula. (Right) Hilbert transform of $Re[Z_L]$ and the measured $Im[Z_L]$.  }
   \label{fig:Fig8}
\end{figure}

Results calculated by the lumped element formula and the standard log formula
are shown in Figs.\ref{fig:Fig6} and \ref{fig:Fig7}.
The lumped element formula may fit the short acceleration gap size of 30 mm
rather than two log formulas which fit longer DUT. 
But the system here consists of four gaps distributed in two meters.
In the case of shorted gaps at an upper part, 
$Re[Z_L]$ by the lumped element formula goes negative in some frequencies, 
which is unphysical (Fig.\ref{fig:Fig8}). 
Moreover Hilbert transform of $Re[Z_L]$ and the measured $Im[Z_L]$ are very far apart.    
This difficulty may come from inappropriate use of the formulas, Eqs.~\ref{eq:5} to \ref{eq:7}. 

\section{$Z_L$ of the FX Kicker System}
\begin{figure}[!htb]
   \centering
   \includegraphics*[width=.9\columnwidth]{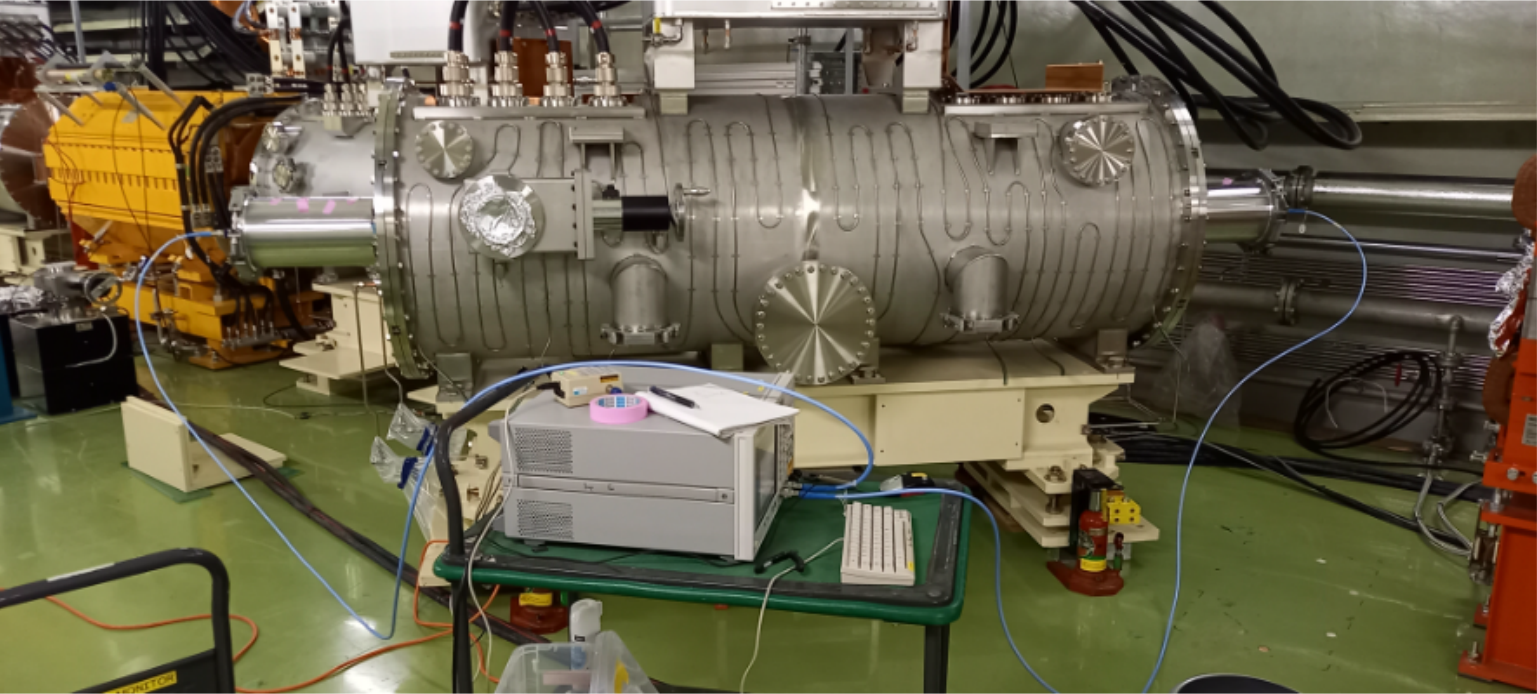}
   \caption{Impedance measurement of the FX Kicker with feeder cables.}
   \label{fig:Fig9}
\end{figure}

\begin{figure}[!htb]
   \centering
   \includegraphics*[width=.52\columnwidth]{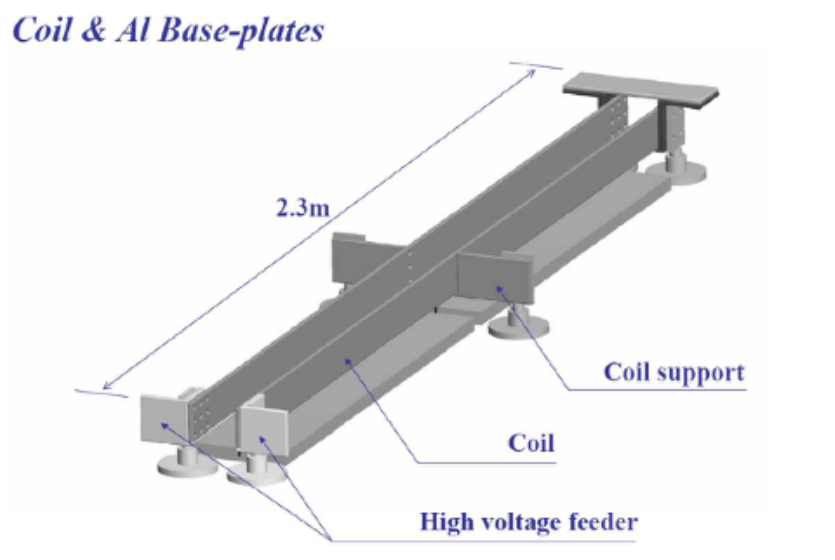}
   \includegraphics*[width=.45\columnwidth]{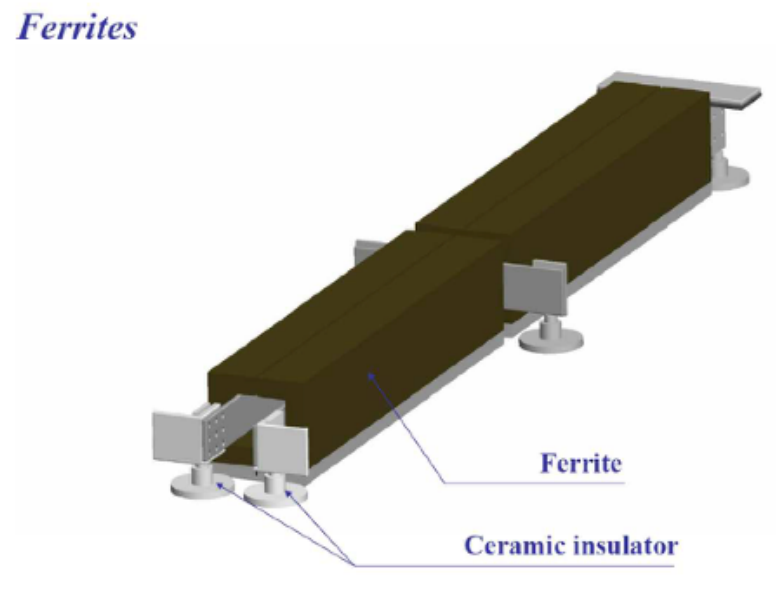}
   \caption{Structure of the kicker.}
   \label{fig:Fig10}
\end{figure}

\begin{figure}[!htb]
   \centering
   \includegraphics*[width=.49\columnwidth]{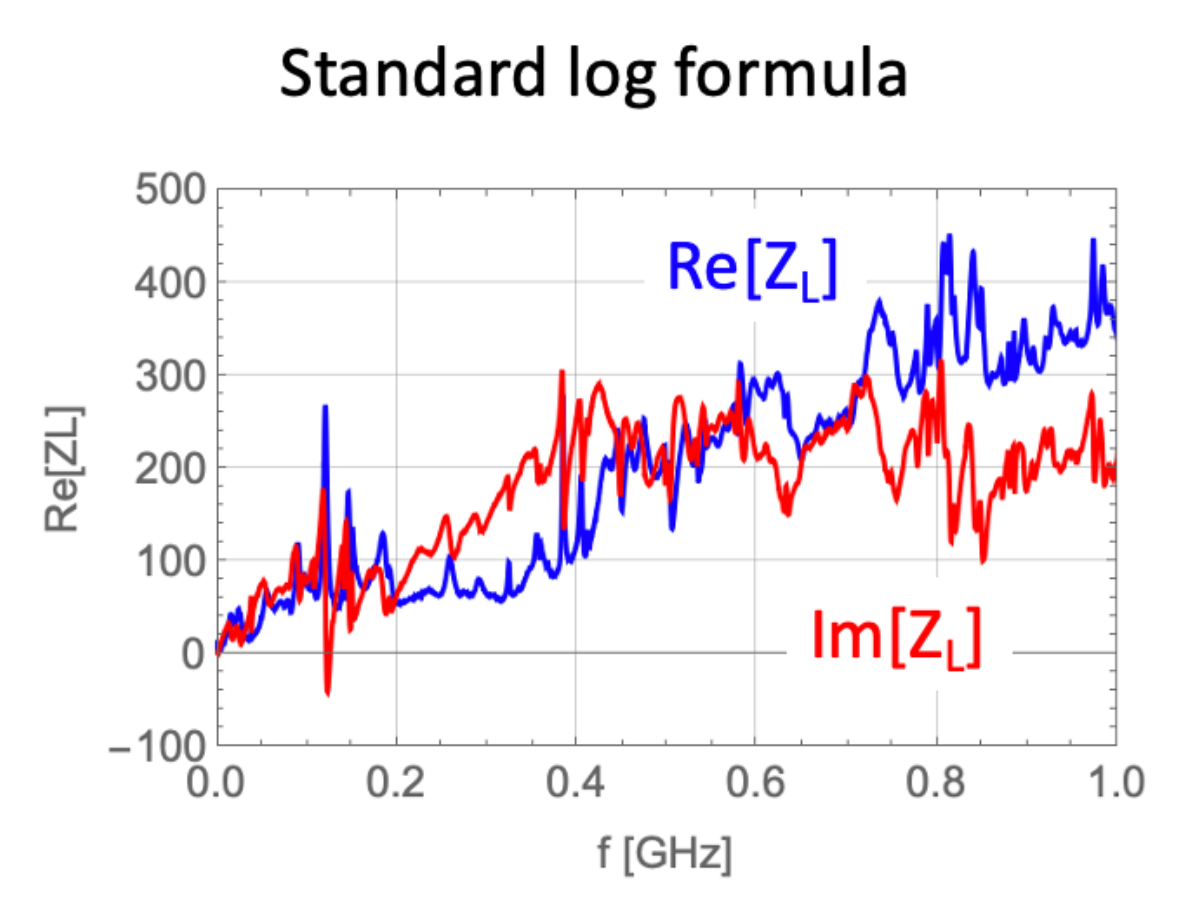}
   \includegraphics*[width=.49\columnwidth]{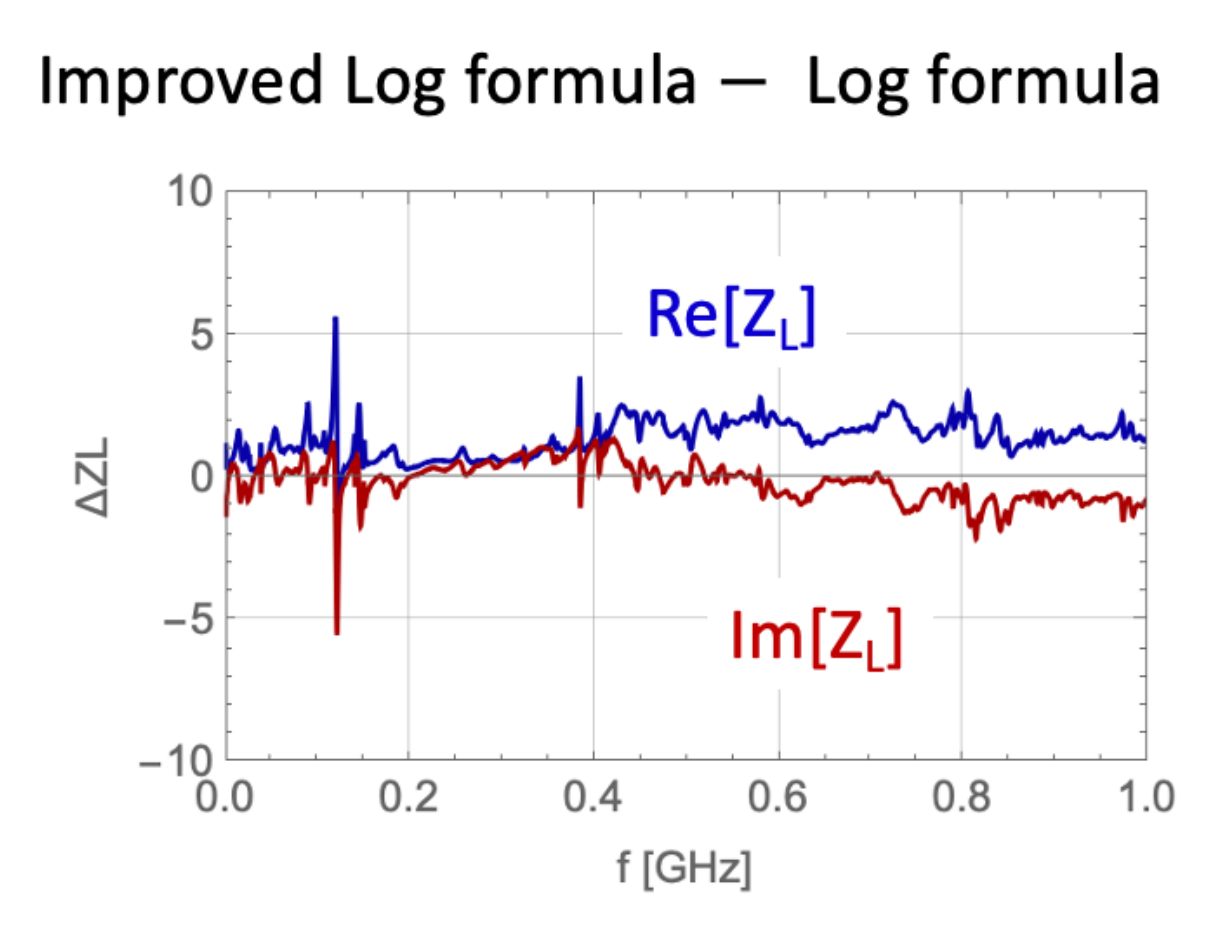}
   \caption{Impedance of one FX kicker. Left figure is calculated with the standard log formula.
   The right is the difference between the impedance calculated by the standard log formula and one by the improved log formula. }
   \label{fig:Fig11}
\end{figure}

There are five tanks for FX kickers at INS-C in the MR \cite{Koseki}.
We measured the fifth tank as shown in Fig. \ref{fig:Fig9} in Jan. and Feb. 2022.
The kicker structure is shown in Fig. \ref{fig:Fig10}. 
A matching circuit is attached on top of the tank. 
The power supply at the power supply building on the ground is connected with a feeder cables. The circuit is almost decoupled by a satuarable inductor. 
The left of Fig. \ref{fig:Fig11} shows the $Z_L$ calculated with the standard log formula.
The difference between the impedance calculated by the standard log formula and one by the improved log formula is a few \% of the $Z_L$ as shown in the right. 

\section{SUMMARY AND PROSPECT}

DUT S-parameters were successfully measured with "2X-THRU de-embedding".
Then longitudinal impedances, $ {Z}_{L} $, are calculated with conventional formulas.
Validity of the process of mapping from $ S_{21} $ to $ Z_L $ has been investigated.  
Hilbert transform of $ Im[Z_L] $ has been tested 
and one pair of data shows large discrepancy.
Inappropriate use may be a reason.
Careful comparison with the CST simulation is also under way.
Besides above reported, transverse impedance measurements with two wires are also performed. Additional measurements with reference pipes are planned.
With the impedances above measured, previously measured and calculated with CST studio, beam stability is being investigated with computer programs.

\ifboolexpr{bool{jacowbiblatex}}%
	{\printbibliography}%
	{%
	
	
} 
%
%
\newpage

\include{annexes-A4}

\end{document}